# Inequality and Concentration in Farmland Production and Size: Regional Analysis for the European Union from 2010 to 2020[1]

**Simone Boccaletti[2], Paolo Maranzano[3]\*, Miguel Viegas[4]**

## 1. Introduction

Understanding the dynamics of industry concentration in the agricultural sector is crucial for policymakers to determine where and how policies aimed at supporting small-scale farms might be most effective. As a matter of fact, the agricultural sector is one of the key sectors that has the ability help the withstand and recover from the economic downturn impact, especially in rural areas (e.g., see Giannakis & Bruggeman, 2018).

At worldwide level, the picture is very fragmented and, consequently, still opaque. Lowder et al. (2021) provide an overview of the number of farms by size at global scale, pointing out that small farms (less than 5 hectares) represent the vast majority of firms, but have less than 20% of the overall farmland. Giller et al. (2021) provide a similar picture, although they recognize that, since production costs and selling prices are determined by large-scale markets, the danger in the next decades is the increase in the marginalization of smallholder farmers, and additional and excessive dependence on very large farms. As of today, the picture of the market is the same as at the beginning of the millennium: the distance described in Von Braun (2005) between the "marginal" farm (small, with low level of sustainability and disconnected to science) and the "dominant" farm (large, more sustainable and users of advanced science) has not closed.

[2] Department of Economics, Management and Statistics, University of Milano-Bicocca, Milano, Italy & Osservatorio O-Fire, University of Milano-Bicocca, Milano, Italy; ORCID: https://orcid.org/0000-0002-6972-5005

[3] Department of Economics, Management and Statistics, University of Milano-Bicocca, Milano, Italy & Fondazione Eni Enrico Mattei, Milano, Italy; ORCID: https://orcid.org/0000-0002-9228-2759

[4] Research Unit on Governance, Competitiveness and Public Policies (GOVCOPP), University of Aveiro, Portugal, ORCID: https://orcid.org/0000-0003-1390-4992

\* Corresponding author at paolo.maranzano@unimib.it

When it comes to the European agricultural sector, two large evidences arise. First, there is a considerable variability across and within region and countries of farms and companies' structures (Guarín et al., 2020). We find in the literature several historical, cultural, geographical, economic, and political factors that explain these differences (Zimmermann et al., 2009). Second, throughout the decades, the European Union's agricultural sector has undergone a significant transformation, marked by a steady concentration of agricultural holdings. This concentration can increase efficiency. Larger farms can potentially achieve economies of scale, leading to lower production costs and potentially higher overall output. However, intensive agriculture practices associated with larger farms can raise concerns about soil degradation, water pollution, and biodiversity loss (Fassò et al., 2023). Moreover, the decline of small farms can negatively impact rural communities, leading to job losses and a weakening of the social fabric.

According to Eurostat estimates, the total number of farms in Europe went from about 12 million in 2010 to just over 9 million in 2020, while standard output increased from 304 billion to nearly 360 billion over the same period (Eurostat, 2024a). Figure 1 clearly illustrates the concentration of the agricultural sectors within the European Union. Between 1990 and 2020, the total number of farms has decreased significantly in most EU countries (we only show countries with data available for both years). It is noticeable that the average area per farm has also increased substantially.

**Figure 1: Land concentration of farms within the European Union**

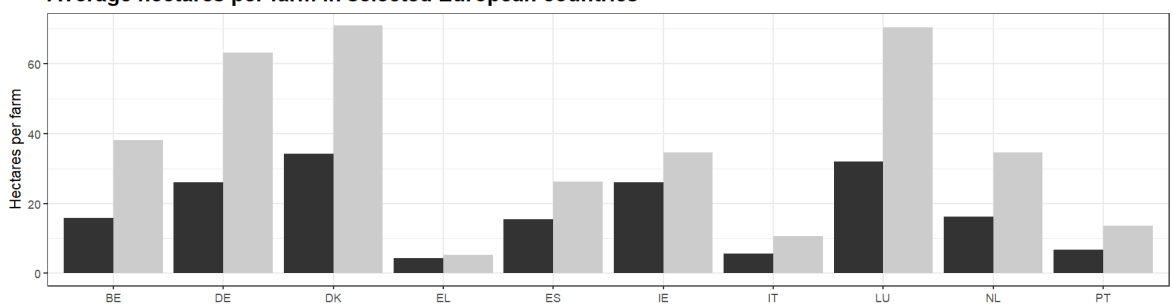

Average hectares per farm in selected European countries

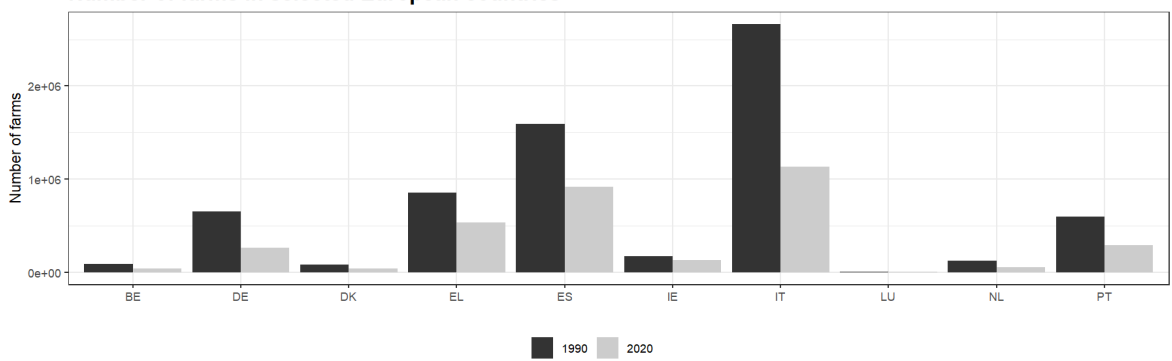

Number of farms in selected European countries

Several factors contributed to this trend. Global competition and volatile market prices push smaller farms towards consolidation or closure. Larger farms can often afford advanced machinery and automation, leading to increased efficiency and productivity. Younger generations are less likely to pursue careers in agriculture, leading to a decline in the available workforce for smaller farms. Lastly, EU Policies, while not directly aimed at concentration, affect land distribution and might inadvertently favor larger farms or smaller ones. Such evidence, legitimately leads to questions about how the structure (e.g., the type of farming and the average size) of farms has changed and whether this change was uniform or heterogeneous within Europe.

In this paper, we investigate the phenomenon of market concentration in the European agricultural and livestock farming industry from 2010 to 2020. We use regional (NUTS-2) and national (NUTS-0) level data from Eurostat (Eurostat, 2024b) to study the spatio-temporal dynamics of disparity and volatility of the main production and economic indicators of European farms. In particular, we are interested in studying the variability within-and-between regions with regard to economic and production size, to assess if the European agricultural market suffered from an increasingly concentration of power in fewer but larger farm holding. We proxy market concentration and disparity by using the Gini Index (Giorgi & Gigliarano, 2017). The general aim of this paper is declined into two research questions:

1) Is there is a common trajectory between (i.e., at the national level) and within (i.e., at the subnational level) European countries of farmland concentration and production concentration between 2010 and 2020?
2) Is there a positive association between land concentration and production concentration?

Our empirical findings confirm the fragmented picture of the European agricultural sector. First, we detect a high level of within-and-between countries heterogeneity for both land concentration and standard output concentration. Also, we detect limited cross-country areas with similar patterns of land and output concentrations. Second, we show the existence of a positive association between the two concentration measures, potentially implying that both economies of scale and economies of scope are relevant in the agricultural sector. Finally, by pursuing a historical outlook, we discuss which choices on the Common Agricultural Policy have been undertaken by the EU authorities and the impacts they have generated on the concentration process evidenced by the data.

To the best of our knowledge, the literature on territorial disparity of the European Agricultural sector is still under development. One of the first notable papers related to our analysis is the one by Vollrath (2007), in which the author show how land inequality is inversely related to productivity. In other words, a decrease in the Gini index of land distribution substantially increases land productivity. This is because farms operated with family labor benefit from productivity advantages, and a more equal land distribution equalizes the marginal product of labor across farms. With this regard, the existence of a sizeable effect of land inequality on output is a symptom of economic inefficiencies in the agricultural sector. The extensive mapping provided by this study contributes to this literature by allowing a fine spatial-scale socio-economic assessment of the European agricultural market integration process, its recent and future trends in the complex and uncertain post-COVID context and the restructuring of international relations due to crises and the green energy transition.

The reminder of this paper is organized as follow. In Section "An historical perspective on the market concentration process in the European Union" we provide an historical compendium about the most relevant agricultural policy interventions and events occurred since the early stage of the European Union (EU) aiming at contextualizing the empirical findings within a socio-political framework. In particular, our focus is on the historical development of the Common Agricultural Policy acts implemented since the '60s. In Section "Background on heterogeneity and distributional issues in the European agricultural sector" we synthesize some relevant issues raised in the agricultural economics literature and directly related to the topic of market concentration. In particular, we focus on the territorial heterogeneities and the unequal distribution of production, profits and resources among farms and the subsequent economic consequences. In Section "Data and methods: assessing agricultural market concentration in Europe using the Eurostat regional database on agriculture", we discuss the dataset used to perform the analysis on agricultural market concentration and the statistical methods used to analyze spatial patterns of concentration. In particular, we describe the data source and its regionalized structure, as well as the analytical definition of the Gini index computed on both production (used as proxy of the economic size) and farmland (used as proxy of physical size) of European farm holdings. In Section "Empirical results", we discuss in detail the empirical findings provided by the exploratory data analysis conducted on the available European regions from 2010 to 2020. Eventually, Section "Discussion, Future Work, and conclusive remarks" summarizes the contents of the paper in light of several cornerstones in the literature, and provides some potential further research streams that could be explored as a follow-up.

## 2. An historical perspective on the market concentration process in the European Union

Article 39 of the Treaty on the Functioning of the European Union (TFEU) defines the specific objectives of the Common Agricultural Policy (CAP):

to increase agricultural productivity by promoting technical progress and ensuring the optimum use of the factors of production, in particular labor;

a) to ensure a fair standard of living for farmers;
b) to stabilize markets;
c) to ensure the availability of supplies;
d) to ensure reasonable prices for consumers.

These objectives are both economic (a, b, c) and social (b, e) in scope. They are intended to safeguard the interests of producers and consumers. The CAP is divided into two pillars: Pillar I, which establishes direct payments to farmers based on historical entitlements and greening requirements; and Pillar II, devoted to rural development funding for a range of measures, including support for farmers, environmental protection, and rural development. As a matter of fact, the wording of the article and the objectives of the CAP have remained unchanged since the Treaty of Rome came into force in 1957. However, these objectives of the CAP could not all be achieved at the same time. The renewal of agricultural labor is today one of the main objectives of the CAP. However, this represents a break with its initial objectives and tools.

The initial phase of the CAP (early 1960s) was heavily influenced by the necessity of food Security and the strong dependence of US supply. Europe had emerged from the devastation of World War II with a depleted agricultural sector and a vulnerable food supply. The CAP aimed to ensure self-sufficiency and prevent food shortages. In the early 1960s, when the CAP was first implemented, European agriculture lagged significantly behind the United States in productivity. American agriculture had already undergone significant modernization following World War II. Mechanization, use of synthetic fertilizers and pesticides, and improved crop varieties led to much higher yields per acre compared to Europe. American agriculture featured larger, more market-oriented farms with economies of scale. This allowed them to invest in advanced machinery and adopt new technologies more readily. The European agriculture was still recovering from the devastation of the war. Many farms relied on traditional, labor-intensive practices, resulting in lower productivity. European agriculture was dominated by smaller family farms with limited resources. This fragmented structure hindered widespread adoption of modern techniques (Viegas, 2021).

Being mainly guided by productivist and protectionist ideology, the first CAP measures were aimed at increasing productivity, accelerating mechanization and promoting a qualified workforce (Garzon, 2006). Thus, In the first decades of implementation of the PAC the tendency was for the agricultural area to be concentrated in increasingly larger properties. The Mansholt Plan, formally known as the "Memorandum on the Reform of Agriculture in the EEC" (European Economic Community) put forward by Sicco Mansholt, the European Commissioner for Agriculture, was launched in 1968

(Stead, 2007). The plan encouraged farm consolidation, with smaller farms merging to create larger, more efficient units. Mansholt advocated for increased investment in research and development to enhance agricultural productivity. The Mansholt Plan would lead to the departure of 5 million farmers from the EU-6 and the redistribution of the land then freed up to increase the surface area of the remaining farms, allowing them to modernize and making motorized mechanization more profitable. Of course, the plan recognized the potential social impact of farm consolidation and proposed measures to support farmers leaving the agricultural sector and finding alternative employment. While not fully implemented, the Mansholt Plan sparked a crucial debate about the future of European agriculture. Even if the ambitions were subsequently revised downwards in the face of negative reactions from the agricultural profession, the plan finally implemented concerned modernization, cessation of activity and training. In fact, between 1966 and 1987 the agricultural workforce (farmers and employees) fell from 3 million to 1.4 million Annual Work Unit (AWU) in France, and from 2.3 million to 0.8 million AWU in Germany. European economies must then face the integration of this workforce into other sectors of activity (Détang-Dessendre et al., 2021). The agricultural surface area has been kept relatively constant over time. On the other hand, the number of farmers and agricultural explorations seemed to decrease. Therefore, average area per farm has grown steadily in a trend so fare (Guiomar et al., 2018).

The Mansholt's model began to show its limitations in the 1980s. Intensive production on increasingly large farms raised concerns about environmental sustainability, but also about social and territorial cohesion (Knickel et al., 2018). The first reform of 1992 (i.e., the so-called Mc Sharry Reform) and later at the turn of the century the "Agenda 2000" with the creation of the second pillar of the CAP dedicated to rural development tried to mitigate this trend, albeit with differentiated impacts in the various member states. Indeed, the second pillar, being co-financed by the Member States, introduced a national component and induced greater heterogeneity within the European Union (Lowe et al., 2002). The measures contained in the second pillar can also help to curb the trend of production concentration. Agri-environmental measures remunerate sustainable practices in disadvantaged areas, contributing to the resilience of these territories (Uthes & Matzdorf, 2013). Aid for the installation of young farmers is also a means of promoting the rejuvenation of farmers and the maintenance of small and medium-sized farms with their generational transmission.

According to Viaggi (2008), maintaining the CAP rules would have led to the continuation of the concentration process of production factors and the disappearance of 20% of farms between 2010 and 2020. This scenario did not materialize, with a reduction that remained at 15%. Between 2010 and 2020, various measures were introduced in the various CAP regulations that may explain this discrepancy. In the first pillar, direct aid decoupled from production (paid per hectare) tends to favor large farms and land concentration (Garrone et al., 2019). On the other hand, coupled payments (paid according to production, especially in ruminants) are crucial for ensuring the viability of small and medium-sized farms (Chatellier & Guyomard, 2023). Moreover, in order to ensure greater equity in the distribution of support, various instruments have been introduced to favor smaller farms. The redistributive payment increases the payment for the first hectares. Modulation and capping aim to reduce the amounts in whole payments or in part from a certain threshold. These are measures that seek to stop the abandonment of the activity by small and medium-sized farmers and mitigate the trend of concentration of agricultural activity. However, these instruments are applied differently in Member States. Many countries do not apply capping at all. As for the redistributive payment, some countries value the first 5 hectares while others value the first 52 (as is the case in France).

In synthesis, while Pillar I has been criticized for potentially exacerbating farm concentration, the impact of Pillar II on farm concentration is more complex and nuanced. Many Pillar II measures are designed to support small and medium-sized farms, potentially mitigating the concentration of land ownership. These measures include young farmer schemes, rural development programs or agri-environmental schemes. By promoting diversification into non-agricultural activities, Pillar II can strengthen the viability of small farms, reducing the pressure to expand or sell land. Certain Pillar II measures, such as agri-environmental schemes, may favor smaller farms with more diversified land use practices. However, other Pillar II measures, particularly those related to investment in physical assets, could potentially benefit larger farms with greater financial resources making impact of Pillar II on farm concentration not so straightforward. In short, while the CAP's Pillar II has the potential to mitigate farm concentration, its effectiveness heavily depends on the specific design and implementation of individual (i.e., at the national level) measures (Henke et al., 2018), resulting in a tendency to increase heterogeneity within the European Union. A careful evaluation of the different measures and their interactions is therefore necessary to assess the overall impact on farm structure.

Concluding, the CAP and national agricultural policies have shaped farm structures through subsidies, market regulations, and land use restrictions. Indeed, the strong tendency of concentration that characterizes the evolution of the agricultural sector from the second half of the 20th century is the result of a massive transformation of the sector, initially desired and supported by both the CAP and national policies (Détang-Dessendre et al., 2021).

# 3. Background on heterogeneity and distributional issues in the European agricultural sector

Distributional issues in the agricultural sector refer to the economic consequences of the presence of farms and companies with different levels of capital endowments (land), productivity level and growth patterns on different economic outcomes, such as health indicators (e.g. food security), economic growth indicators (e.g., GDP growth) and inequality indicators (e.g., income inequality). As a matter of fact, in the last two decades there has been an increased attention on the disparities in land ownership, with a heightened awareness of how these inequalities contribute to broader economic and social divides. Moreover, the discussion on land disparity calls also for a concurrent analysis of productivity disparity of the agricultural sector, in terms of both land productivity and also in terms of labor productivity.

The recognition of disparity among farms has been a well-established concept since the mid-20th century. However, the debate surrounding the types and scales of agricultural systems, as well as the distribution of farms based on size, has gained significant momentum since the early 2000s. This intensification has coincided with the growing global focus on issues such as world hunger, inequality in land distribution, and the need to address the challenges faced by rural communities and extremely poor areas. The ongoing debate is not just about the productivity of different farming systems but also about their social and economic implications. Discussions now frequently focus on finding the balance between large-scale, industrial agriculture and smaller, often more sustainable farming practices that might better support equitable development and reduce poverty in rural communities. This shift reflects a growing consensus that addressing inequality in land distribution and supporting diverse agricultural systems are crucial steps towards achieving global food security and reducing poverty, as well as the sustainability goals by different authorities.

As previously mentioned, the structure of agricultural land in the European Union varies significantly within and between countries. Historical land ownership patterns, such as feudalism or land reforms, have shaped the size and distribution of farms as well as post-war agrarian reforms in some countries that led to land redistribution and affected farm structures. Poland, Czechoslovakia, and East Germany implemented land reforms as part of their communist transitions, breaking up large estates and distributing land to peasants (Mathijs, 2018). While less radical than in Eastern Europe, Italy also underwent land reforms, particularly in the south, to address issues of land ownership and agricultural productivity (Bonanno, 1988).

Different climatic conditions and terrain also influence the types of crops and livestock suitable for specific regions, impacting farm size and specialization. Variations in soil fertility and quality affect agricultural productivity and farm size. The amount of available agricultural land per capita varies across countries, influencing farm size and structure.

On the socio-economic side, several factors can be called upon. First and foremost, the variability depends on the specialization of the different areas and/or the coexistence of crops and livestock in the same area. Second, there is a large gap between rural communities and larger-size farms, not only in the scale of production but also under other aspects such as employment and social inclusion. However, the role of large-scale investment in rural areas is still under debate (Nolte & Ostermeier, 2017). While this type of investments is recognized as an important catalyst for development in rural areas, some authors suggest that they could potentially alter the rural community's landscape. Also differences in agricultural productivity and technology adoption can influence farm size and competitiveness. The demand for specific agricultural products can drive farm specialization and size (Angus et al., 2009). Other Factors like urbanization and environmental regulation can also influence land size structure. The expansion of urban areas can reduce the amount of available agricultural land and influence farm structures (Fischel, 1982). Restrictions on land use and agricultural practices can impact farm size and management. Eventually, different market forces and macro-shocks could impact different areas in many different ways. For example, the invasion of Ukraine by Russia had a relatively large impact on the agricultural sector due to the important role of the two countries. Specifically, the invasion caused a shift in international trade countries and companies had to seek alternative sources for their agricultural imports, which often involved higher costs and longer supply chains. Some countries to reassess their food security strategies, including boosting domestic production or seeking new trading partners.

The distribution of agricultural land is also affected by other factors associated with the distribution of wealth (Stiglitz, 1969). Although J. Stiglitz, in his seminal paper focused mainly on income and wealth, the underlying principles can be extended to analyze the concentration or dispersion of agricultural land ownership. Inheritance laws can significantly influence the distribution of agricultural land. If inheritance laws favor wealthy landowners, it can lead to a concentration of land ownership in fewer hands. Conversely, laws promoting equal division of land among heirs can result in smaller, fragmented landholdings (Huning & Wahl, 2021). Imperfections in the land market, such as information asymmetries, transaction costs, and limited access to credit, can hinder the efficient transfer of land and contribute to land concentration. Moreover, in agriculture, economies of scale often favor larger farms, leading to a concentration of land ownership. However, market imperfections can prevent small farms from achieving optimal size. Government policies, such as land redistribution programs, can significantly alter the distribution of agricultural land as well as subsidy programs that impact land ownership by favoring certain farm sizes or types of production.

Finally, land taxation policies and land use regulations or property rights can also influence land ownership patterns, with higher taxes potentially discouraging land hoarding.

These factors have interacted over time to create the diverse agricultural land structures observed across the EU. Some countries have predominantly small family farms, while others have larger, more specialized farms. We can also observe within some European states various types of agricultural property structures. The case of France is interesting, because it allows us to tie the political variable (common to the entire territory) and observe the effect of the remaining geographical, cultural and historical factors (Landais, 1996). France presents a complex and diverse agricultural landscape, shaped by a combination of historical, geographical, economic, and social factors. The French agricultural sector is primarily composed of family farms, with a significant number of small-scale holdings. However, there is a significative regional diversity. France's agricultural land structure is deeply rooted in its historical land ownership patterns, including the French Revolution's impact on land redistribution (Finley et al., 2021; Rosenthal, 1990). However, the expression of these social and political upheavals varied according to the geographic and climatic characteristics of each region. Indeed, the country's varied topography and climate have led to a mosaic of agricultural systems, from the intensive farming of the Paris Basin to the extensive livestock production in the Massif Central. As in most European states, the decline in rural populations has contributed to the consolidation of farms in some areas, while small-scale agriculture persists in others.

The disparity between the number of holdings and land distribution is shown by different report by FAO (e.g., see FAO, 2014; FAO, 2021) and different academic papers (e.g., see Deininger & Byerlee, 2012; Lowder et al., 2021). The 2021 report by FAO (FAO, 2021) provides the results of a census run between 2006 and 2015 at worldwide level (and, for most of European countries, run earlier than 2012) and provides a picture of the trend at country and aggregate level from the first available data in 1930 (though for a limited number of countries). One of the major things that emerges was the different trend of the European continent with respect to other areas: as at the global level, the trend in average land size of holding decreased from 1960 to 2000 and then increased in the subsequent decade, in Europe the trend in the same period is overall increasing (with the only exception being the 2000 level, slightly lower than the one in 1990). On the contrary, in America and in Oceania, between 2000 and 2010 there was a substantial drop on the size of holdings, while Asia and Africa follow similar patterns. In Europe, the average increase in the size of holdings is common to most of the countries, but with very different magnitudes: while for numerous of countries (e.g., Germany, Denmark, Belgium, Luxemburg, Slovenia and Spain) the size of the holdings in 2010 doubled (or almost doubled) the 1990 level, for other countries this increase is limited (e.g. for Italy, in the same period, the average size of holdings went from 7.5 ha to 10.5 ha; Poland went from 8.3 to 11.3;

Finland recorded 61.9 ha in 1990 and 97.9 in 2010). This heterogeneity has to be considered in conjunction with the fact that, while small farms are the vast majority of the farm operating in the agricultural sector, they operate with limited agricultural land and produce a small amount of food: in Europe, the share amount of food produced from smallholders is around 8% (Lowder et al., 2021).

The observed heterogeneity across European countries may also reflect deeper disparities within individual nations. Historically, regions within the same country, such as Northern versus Southern Italy or Eastern versus Western Germany, have followed distinct development trajectories, leading to significant internal differences. To the best of our knowledge, the study of land disparity and its evolution within Europe and within European countries is still under development, also because the agricultural market is becoming increasingly dynamic[5]. In the latest years, some academic works have been published on related issues and using more granular databases (i.e., considering areas at NUTS-2 level), but with different focuses. As an example, Schiavina et al. (2022) provided a picture of the expansions of Built-up areas with the primary focus on land consumption and its relation to demographic trends. Tóth (2023) uses the concept of territorial capital, which includes considerations on land usage, artificial land, croplands and woodlands, to maps regions at the NUTS-2 level, to understand territorial capital endowments. Both of those studies clearly show intra-region differences, as well as a marked difference between eastern European regions and western European regions.

If land distribution represents the major disparity issue in the agricultural sector, the second relevant level of disparity regards productivity, in various declination: land productivity, labor productivity, capital productivity, total factor productivity, and so on. Agricultural productivity differential were key issues at the European level since the early 2000, in the context of EU enlargements and country adaptation to the Common Agricultural Policy, especially after 2013. Spicka (2013) shows the differential between EU-12 and EU-27, on different economic variable related to the agricultural sector (such as labor demand, compensation of employee and production), showing large gaps between countries in terms of development and means of production. Kijek et al. (2019) while confirming the result on the differential of productivity, show some evidence of convergence, meaning that lower productivity countries effectively put some effort in reducing their productivity gap.

Productivity levels are negatively associated with land disparity. Vollrath (2007) shows this relationship at country level, considering the output per hectare as the productivity variable and the Gini coefficient for land holdings for land disparity, using data retrievable from the FAO world census. The paper also estimates the magnitude of this association: a 0.16 drop in the Gini coefficient[6] at country level is associated to an increase in productivity by 8.5%. These results suggest that an unequal distribution of land is inefficient and can lead to significant negative effects at aggregate level. In other words, when land is unevenly distributed, it often means that resources are not being used in the most productive way. This can result in lower agricultural output and inefficient use of scarce resources.

---

[5] As an example, in recent years, there has been a need to adapt agricultural production to more sustainable practices. Notably, the banning of certain chemical products has had a profound impact on many crops, forcing a re-assessment of agricultural practices and challenging farmers to find alternatives that are both effective and environmentally friendly.

While Vollrath (2007) considered data at country level, Ezcurra et al. (2008) analyzed data from 1980 to 2001 on European NUTS 2 regions, to analyzes whether there is presence of spatial dependence in the regional distribution of productivity in the agricultural sector. In this paper authors use the e Moran's I-statistics using a weighted matrix based on the 10-nearest neighbors, calculated using the geographical distance between the corresponding regional centroids. Their results, while showing that heterogeneity in regions' level of productivity is relevant, adjacent regions tend to be characterized by similar levels of agricultural productivity. They also show that spatial disparity of productivity remains constant throughout the time window considered.

More recently, Giannakis and Bruggeman (2018) focus on the main determinants of labor productivity considering spatial (regional) differences between 2007 and 2013, with the aim to classify European regions based on the agricultural systems' standard output per annual work unit. With a cluster analysis based on different type of variables (environmental, structural, technical and macro variables), authors are able to classify European regions at the NUTS-2 level for six agricultural system (field crops, horticulture, permanent crops, grazing livestock, granivores and mixed crop-livestock) between high and low labor productivity. They show high heterogeneity and high differences between regions, but highly dependent on the agricultural system considered For example, for field crops, there is a somehow clear separation between eastern and central-western Europe whereas for permanent crops, high labor productivity regions are concentrated in the central Europe. Other papers that focused on productivity have narrower scope, considering only specific countries or propose a static approach (e.g., see Lazíková et al., 2021; Martínez-Victoria et al., 2019; Martínez-Victoria et al., 2018; Popescu et al., 2016; Toma et al., 2023).

---

[6] A 0.16 drop is actually a large drop, as we are going to show in this paper: for the totality of countries analysed in this paper, in the years between 2010 and 2020 the absolute variation of the Gini index is lower than this level.

## 4. Data and methods: assessing agricultural market concentration in Europe using the Eurostat regional database on agriculture

We considered regionalized data on the agricultural market in Europe provided by the Eurostat (2023) according to the 2010 Nomenclature of territorial units for statistics classification (NUTS 2010). In particular, we considered regional (NUTS-2) and national (NUTS-0) level information on European farms contained in the "Main farm indicators by NUTS 2 regions" open database (Eurostat, 2024b). The database provides several structural and economic indicators on the agricultural industry in Europe for the years 2010, 2013, 2016, and 2020.

Among the full set of available data, we considered a subsample concerning the following regional quantities:

- Overall number of agricultural holdings or farms (count)
- Utilized agricultural land (measured in hectares)
- Standard output from agricultural production (measured as Euros)

Notice that, given the exploratory purposed of this paper, we decided to do not consider further subclassification of farms with respect to the productive specialization, such as the organic producers or the livestock-specialized farms. However, future researches should take into account such information in order to extend our findings and provide a broader characterization of the market concentrations dynamic in the sector. We considered the largest possible spatio-temporal sample by preserving European regions having complete (non-missing) information for the entire time span. Despite more recent classification are available, we adopted the NUTS 2010 nomenclature as it is the only one fully covering the 27 countries currently belonging to the European Union within the 2010-2020 period. The selected dataset includes complete data for 236 NUTS-2 regions.

Concentration indices for both the farmland (i.e., the utilized agricultural area) and the production (i.e., the standard output from farming) are computed considering the stratification of agricultural farms into $K = 11$ classes[7] of economic size based on increasing standard output values. By taking advantage of the number of farms and the cumulated standard output for each stratum in region, the Gini Indices for production and farmland are computed following Cerqueti et al. (2024), which in turn rely on the Gini index specification for grouped data by Brown (1994). Specifically, let $d$ be the index for the production ($d = P$) and the farmland ($d = L$) and let $j = 1, ..., K$ the index for the economic strata. We then compute the Gini index for each region $s = 1, ..., 236$ as follows:

$$Gini_{ds} = \frac{N_s}{N_s - 1} \left[ 1 - \sum_{j=1}^{K} [(Q_{dj} + Q_{dj-1}) \times (F_{dj} - F_{dj-1})] \right]$$

where $d = \{P, L\}$ identifies the production ($P$) or the farmland ($L$) values; $N_s$ is the total number of farms in region $s$; $Q_{dj} = \sum_{i=1}^{j} q_{di}$ is the cumulative proportion of production or farmland up to the $j$-th ordered class (with $q_{di}$ being the regional share of production or farmland associated with the $i$-th ordered class over the total), and $F_{dj} = \sum_{i=1}^{j} f_{di}$ is the cumulative proportion of farm holding up to the $j$-th ordered class (with $f_{di} = \frac{N_{si}}{N_s}$ being the regional share of farms associated with the $i$-th ordered class over the total number of farms in the region under the constraint $\sum_{i=1}^{K} N_{si} = N_s$). In synthesis, $Gini_{Ps}$ represents the Gini index for the standard output (production) in region $s$, while $Gini_{Ls}$ represents the Gini index for farmland in region $s$. Notice that, as we consider yearly data for 2010, 2013, 2016, and 2020, we computed a Gini index for each year and region.

---

[7] The current Eurostat classification of standard output is as follows: zero euros; over zero euros to less than 2000 euros; from 2 000 to 3 999 euros; from 4 000 to 7 999 euros; from 8 000 to 14 999 euros; from 15 000 to 24 999 euros; from 25 000 to 49 999 euros; from 50 000 to 99 999 euros; from 100 000 to 249 999 euros; from 250 000 to 499 999 euros; 500 000 euros or over.

From the statistical perspective, the Gini index can be interpreted either as a statistical dispersion (i.e., variability) measure or a statistical concentration measure (Giorgi & Gigliarano, 2017). Indeed, while the first index measures the agricultural market concentration with respect to the economic capacity of farms, the latter measures the agricultural market concentration in terms of land owned by farm holding, that is, with respect to the physical size. By definition, the Gini index is a normalized metric which lies between 0 and 1, or, equivalently, between 0 and 100 if rescaled in a percentage scale (Giorgi & Gigliarano, 2017). A value of Gini index equal to 0 is expected when all the farms in a given region have the same standard output or they hold the same hectares on land (i.e., perfect equal distribution scenario). Conversely, a value close to 1 represent a situation of high concentration in which almost all the land or standard output is owned by a very restrict number of farm holdings, leaving only a very small amount to the remaining companies. In the extreme case of the Gini index being exactly 1, the entire agricultural land (standard output) of a region would be owned (produced) by a single farm holding (i.e., full concentration scenario).

In this paper, we aim at describing the spatial and temporal evolution of agrobusiness concentrations by comparing the two Gini indices and establishing an empirical relationship among the concentration of production (used as proxy of the economic size of farms) and the concentration of farmland (used as proxy of physical size of farms). The exploratory analysis is performed by studying the evolution in space and time of the linear correlation between the concentration of agricultural land and the concentration of agricultural production, as well as by investigating the temporal dynamics of spatial autocorrelation measures that can describe the influence of the Gini index recorded in neighboring regions on that observed in each of the European regions.

When addressing the research question on the correlation among concentration measures, although it is reasonable to expect a direct and positive relationship between the two concentration measures, some careful consideration is required. It is straightforward that if the Gini index of farmland is equal to 1 (i.e. one firm owns all the land in a region), this directly implies a standard output Gini index of 1. On the contrary, a Gini index of farmland close to zero does not directly imply a low Gini index of production. Two considerations support this view. Firstly, even if the land is homogeneously distributed, some of it could be 'inactive', leading to an uneven distribution of standard output. Second, even if most of the land is productive, two regions with the same land Gini index could have higher or lower levels of production concentration for different reasons: farms in one region could have higher productivity levels due to differences in land use productivity (e.g. organic vs. non-organic production; crops vs. livestock, etc.), but also due to the cost structure and market characteristics of the farms. Finally, the link between the two levels of inequality depends to a large extent on possible economies of scale in the agricultural sector. A fall in the Gini index for land could either mean that some micro farms are acquiring other small plots of land and becoming larger, or that a large area of land has been sold to a number of other farms. The possible change in the standard output Gini index could be driven by possible economies of scale following this decrease in land concentration. If economies of scale are relevant, a decrease in land concentration could lead to a decrease in the heterogeneity of standard output as farms become more similar in terms of physical size.

Eventually, for what concerns the spatial dependence analysis, we recall that as clearly remarked by Ezcurra et al. (2008), the European agricultural market shows strong evidence of local patterns in the spatial distribution (both in terms of heterogeneity and spatial dependence) that result in the coexistence of sub-areas with nonhomogeneous characteristics. Their findings regarding agricultural productivity suggest that there are at least two sub-areas, north-central Europe oriented toward animal farming, with a relatively large share of cereal and forage crops, and southern Europe specializing in the production of vegetables and permanent crops, confirming the hypothesis of dualism in the European agricultural market discussed in Kearney (1991) and Gutierrez (2000). In addition, evidence of spatial dependence among regions (i.e., neighboring regions tend to register similar levels of gross value added per worker in the agricultural sector) is confirmed for a time period of almost 20 years between 1980 to 2001. In the present study we are not interested in measures of agricultural sector productivity; rather, we want to assess the evolution in space and time of the statistical concentration of production and farmland as measured by the Gini Index. In addition, we consider a sample referring to a time frame subsequent to that of Ezcurra et al. (2008) which spans from 2010 to 2020 and covers a larger number of European regions (we also include all regions of the Eastern European Union countries). To make the two studies comparable, we also implement Exploratory Spatial Data Analysis techniques (Elhorst, 2010; LeSage, 2008) designed to measure the degree of spatial dependence of agricultural concentration and its temporal evolution. In particular, we estimate the dependence between regions using Moran's statistic for both global and local spatial autocorrelation (Anselin, 1995). Several studies involving spatial econometrics tools to explore regional inequality patterns (e.g., see Puttanapong et al., 2022; Touitou et al., 2020) our using inequality as a spatial determinant (e.g., see Panzera & Postiglione, 2022) can be found in recent literature.

## 5. Empirical results

In this section, we firstly provide a broader picture related to the analysis at country level, and then we dig deep into the dynamics of the Gini index for both the production and farmland. The main empirical findings regarding both the national and regional spatio-temporal dynamics of agricultural market concentration are summarized in Table 1. In the table, we synthetize the evolution of the phenomenon between 2010 and 2020 by emphasizing the country-specific characteristics with respect to the temporal and the territorial (i.e., intra-country) dynamics. In general, at country level, the analysis shows a pronounced country heterogeneity, for both the farmland Gini index and the production Gini index. In addition, land size is, for most of the countries, less concentrated than the standard output. This implies that overall land productivity is highly heterogeneous between and within countries.

**Table 1: Summary of the main empirical results on the Gini index for farmland and production**

| | Farmland Gini Index | | Production Gini Index | |
|---|---|---|---|---|
| **Country** | **Time variation** | **Intra-country differences** | **Time variation** | **Intra-country differences** |
| **Austria (AT)** | Constant after 2013 | Some differences detected, but province heterogeneity decreases over time | Constant after 2013 | Minor differences detected |
| **Belgium (BE)** | Negligible | Minor differences detected | Negligible | Minor differences detected |
| **Bulgaria (BG)** | Decreasing in the time horizon considered | Minor differences detected | Increasing until 2016, then slightly decreasing between 2016 and 2020 | Minor difference detected, but province heterogeneity decreases over time |
| **Cyprus (CY)** | Increasing until 2016, then slightly decreasing between 2016 and 2020 | | Slight increase over time | |
| **Czech Republic (CZ)** | Constant over time | Minor differences detected | Decreasing until 2013, then increasing | Minor differences detected |
| **Germany (DE)** | Slight increase over time | Some differences detected, especially between East and West Germany | Slight increase over time | Some differences detected, especially between East and West Germany |
| **Denmark (DK)** | Increasing over time | | Slight increase over time | |
| **Estonia (EE)** | Considerable time variations | | Constant until 2016, then slightly decreasing | |
| **Greece (EL)** | Considerable decrease between 2013 and 2016 | Some differences detected | Small time variations detected | Minor differences detected |
| **Spain (ES)** | Slight increase until 2016 | Considerable differences, especially between Northern regions and Southern regions | Constant over time | Minor/negligible differences |
| **Finland (FI)** | Slight increase over time | Minor/negligible differences | Slight increase over time | Minor/negligible differences |
| **France (FR)** | Slightly increasing until 2016, then sharply decreasing | Minor differences detected, but province heterogeneity seems to decrease over time | Slightly decrease over time | Minor differences detected |
| **Croatia (HR)** | Considerable time variations | Considerable differences, but province heterogeneity seems to decrease over time | Increasing over time | Considerable differences detected |
| **Hungary (HU)** | Constant until 2016, then sharply decreasing | Minor/negligible differences | Time variations detected, but very similar 2010 and 2020 levels | Minor/negligible differences |
| **Ireland (IE)** | Small time variations detected | Minor/negligible differences | Small time variations detected, but very similar 2010 and 2020 levels | Minor/negligible differences |
| **Italy (IT)** | Sharp decrease between 2010 and 2013 | Regional differences detected, but not differentiated between northern and southern Italy | Time variations detected, but very similar 2010 and 2020 levels | Regional differences detected, but not differentiated between northern and southern Italy |
| **Lithuania (LT)** | Slightly increasing over time | | Slightly increasing over time | |
| **Luxembourg (LU)** | Constant over time | | Slightly increasing over time | |
| **Latvia (LV)** | Increasing over time | | Slightly increasing over time | |
| **Malta (MT)** | Decreasing over time | | Decreasing until 2016, then increasing | |
| **Netherlands (NL)** | Small time variations detected; Noticeable drop between 2013 and 2016 | Minor/negligible differences | Decreasing until 2016, then constant | Minor/negligible differences |
| **Poland (PL)** | Constant over time | Considerable differences detected | Slightly increasing over time | Minor differences detected |
| **Portugal (PT)** | Constant over time | Considerable differences detected | Slightly increasing over time | Minor differences detected |
| **Romania (RO)** | Time variations detected | Considerable differences detected | Increasing over time | Considerable differences detected |
| **Sweden (SE)** | Slightly increasing over time | Negligible differences | Slightly increasing over time | Negligible differences |
| **Slovenia (SI)** | Small time variations detected | Negligible differences | Constant until 2016, then slightly increasing | Negligible differences |
| **Slovakia (SK)** | Slightly decreasing over time | Negligible differences | Slightly decreasing over time | Negligible differences |

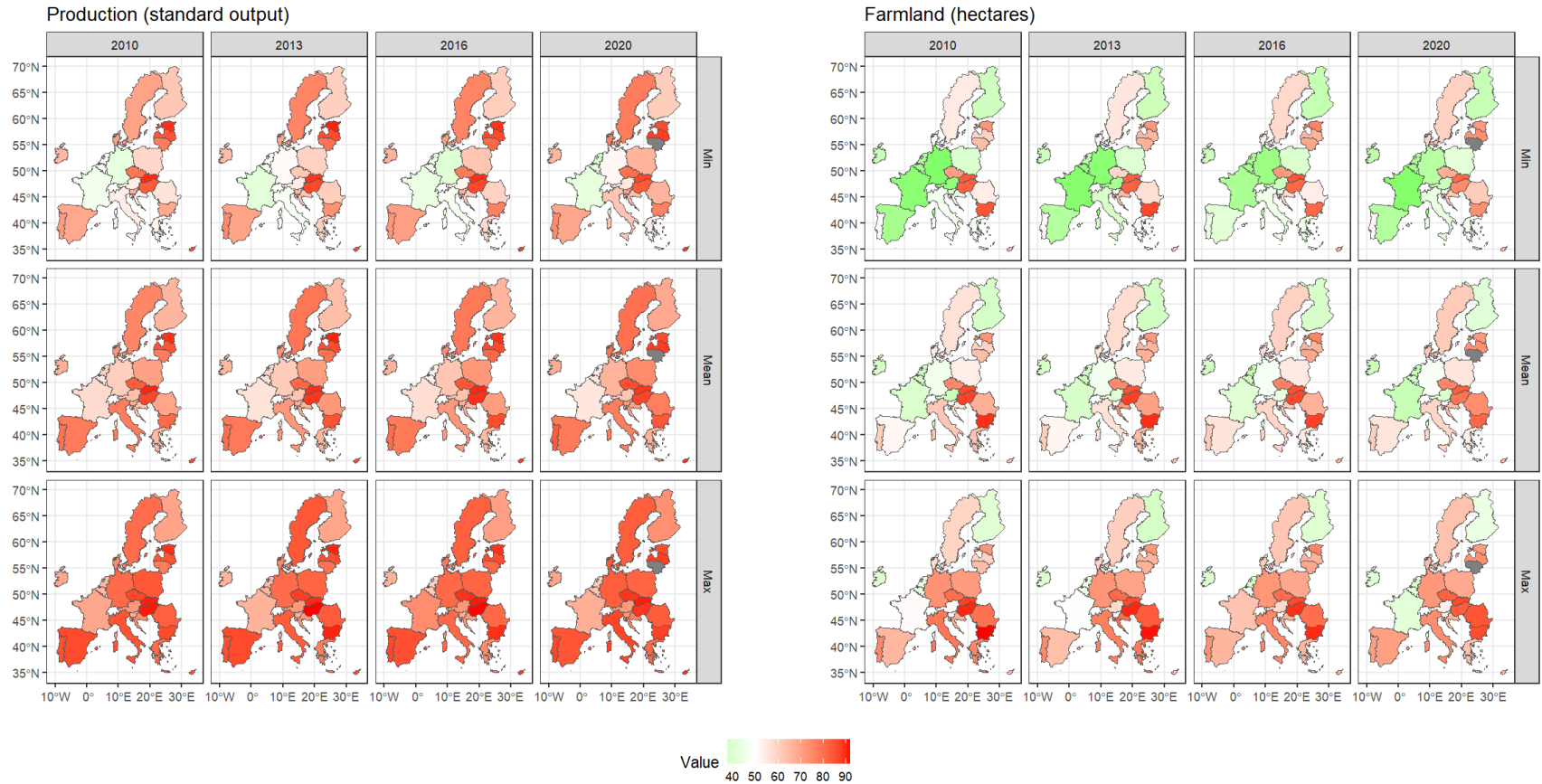

Figure 2: Maps of descriptive stats for Gini index from 2010 to 2020 in Europe

Note: Gini index is reported in a 0-100 scale. By rows are reported the minimum (Min), the average (Mean), and the maximum (Max) Gini index by country and year computed aggregating the available NUTS-2 values. Regarding Lithuania (LT), the last available information from Eurostat regards 2016. Thus, Gini index and the corresponding descriptive statistics for 2020 are not available (grey region).

Figure 2 shows the minimum, the average and the maximum Gini index at the country level (NUTS 0) for both Standard Output (left panel) and agricultural land (right panel). At a first glance, some country specificity clearly emerges, while, on average, it seems that these statistics seems overall constant over the time horizon considered in this paper. For a more detailed insight, in Table A1 of the Appendix we report the numerical value of the average Gini index by country and year for both standard output and hectares, whereas in Table A2 we report the country-and-yearly-specific standard deviations as a rough measure of intra-country variability.

The two Gini indexes take similar values only for Bulgaria, Croatia, Hungary, Romania and Slovakia. Also, countries with relatively low levels of the Gini index are mostly concentrated in the central part of Europe. Specifically, France, Finland, the Netherlands, Belgium and Austria register an average hectares Gini index lower than 40%, with a standard output Gini index that oscillates between 50% and 65%. Even Germany and Slovenia seem to have a relatively homogeneous market concentration.

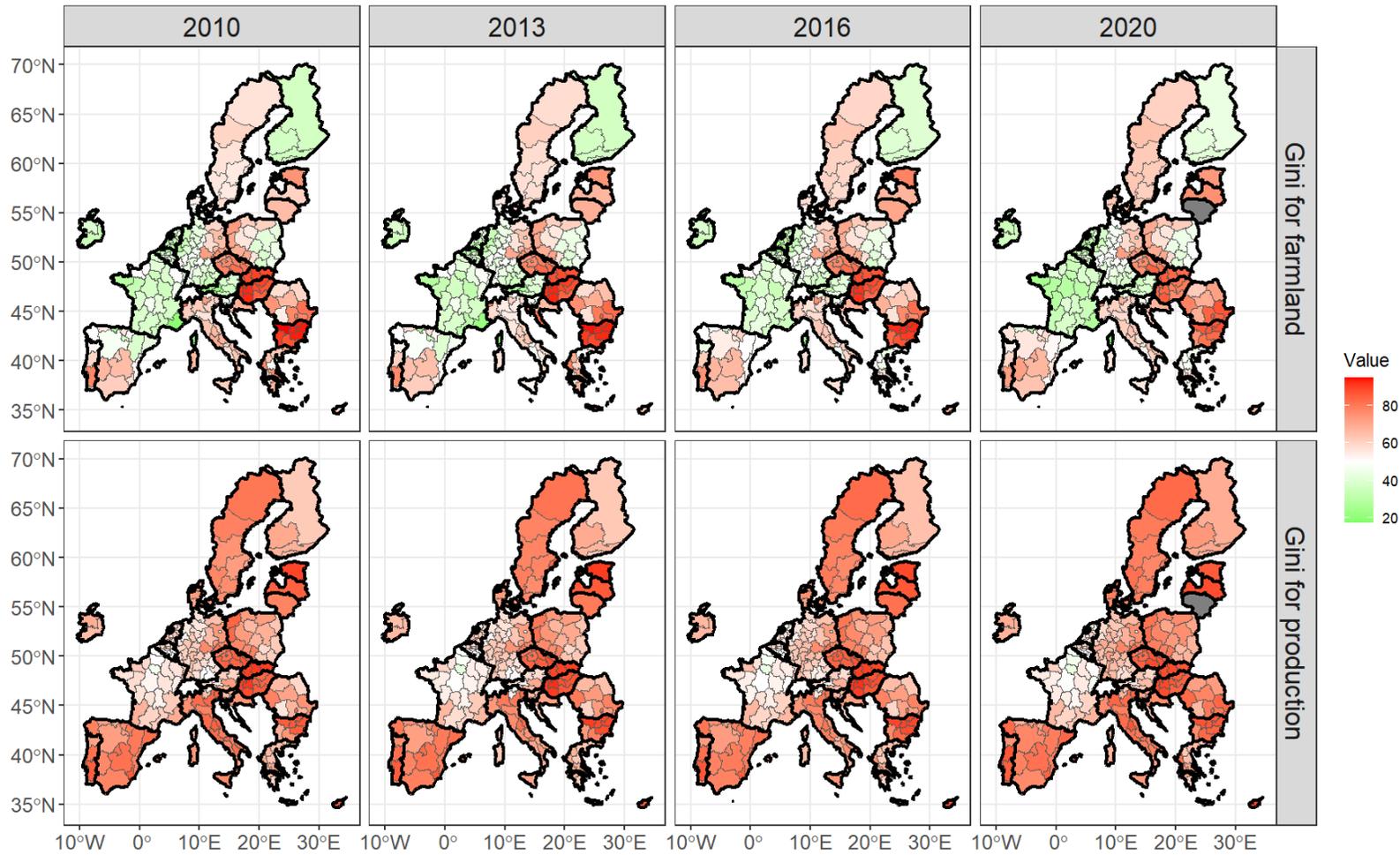

Note: Gini index is reported in a 0-100 scale. Regarding Lithuania (LT), the last available information from Eurostat regards 2016. Thus, Gini index for 2020 is not available (grey region).

Figure 3 shows the Gini index at the regional NUTS-2 level for both the production (second row) and the farmland size (first row) in the four years considered.

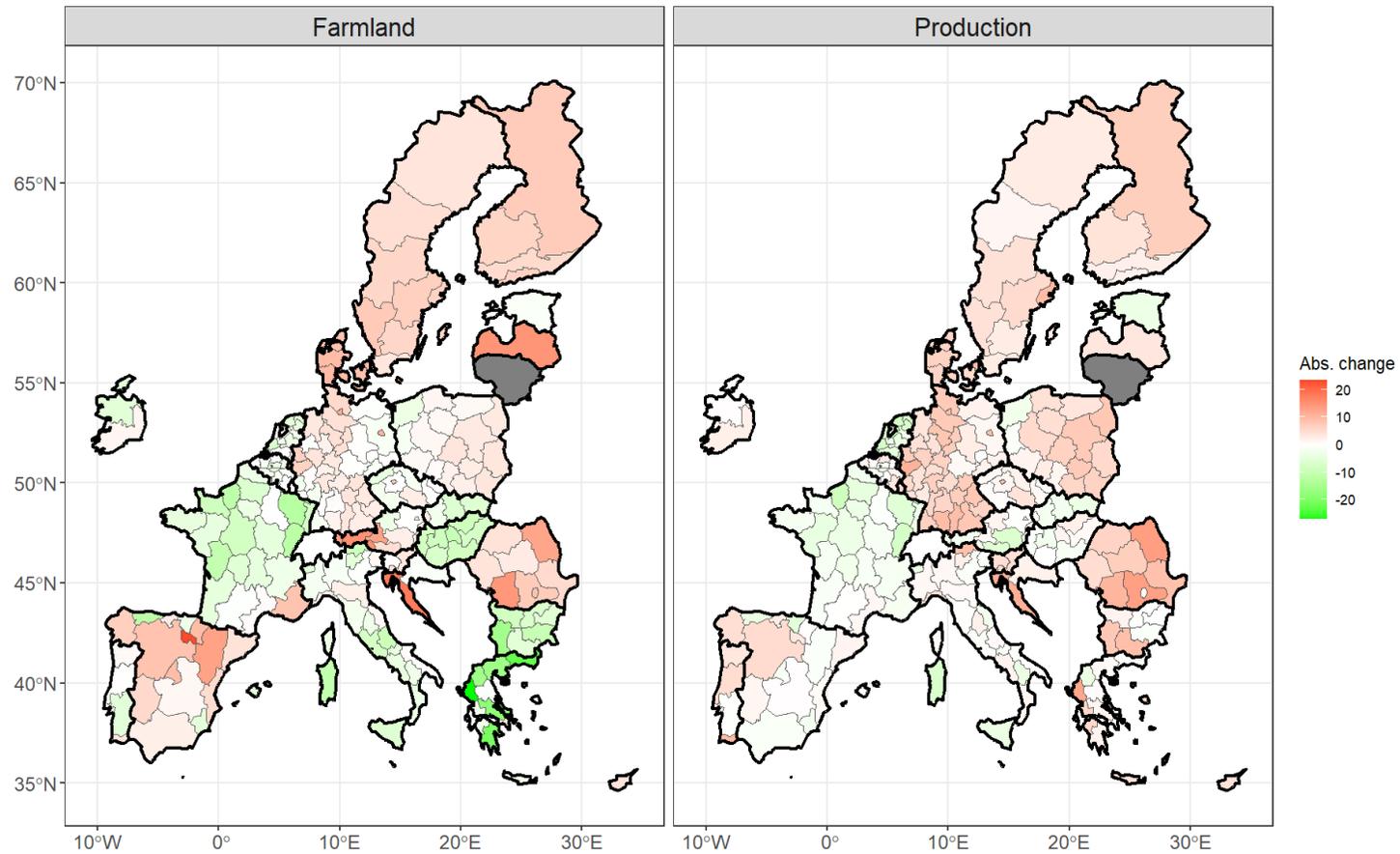

Figure 4: Maps of raw change in Gini index for production and farmland from 2010 to 2020 in Europe

Note: Variations of the Gini index is computed as the difference between the regional (NUTS-2) Gini index for 2020 and the Gini index for 2010. As the Gini index are computed on a 0-100 scale, also the variations are in the same scale. Regarding Lithuania (LT), the last available information from Eurostat regards 2016. Thus, Gini index and the corresponding variation is not available (grey region).

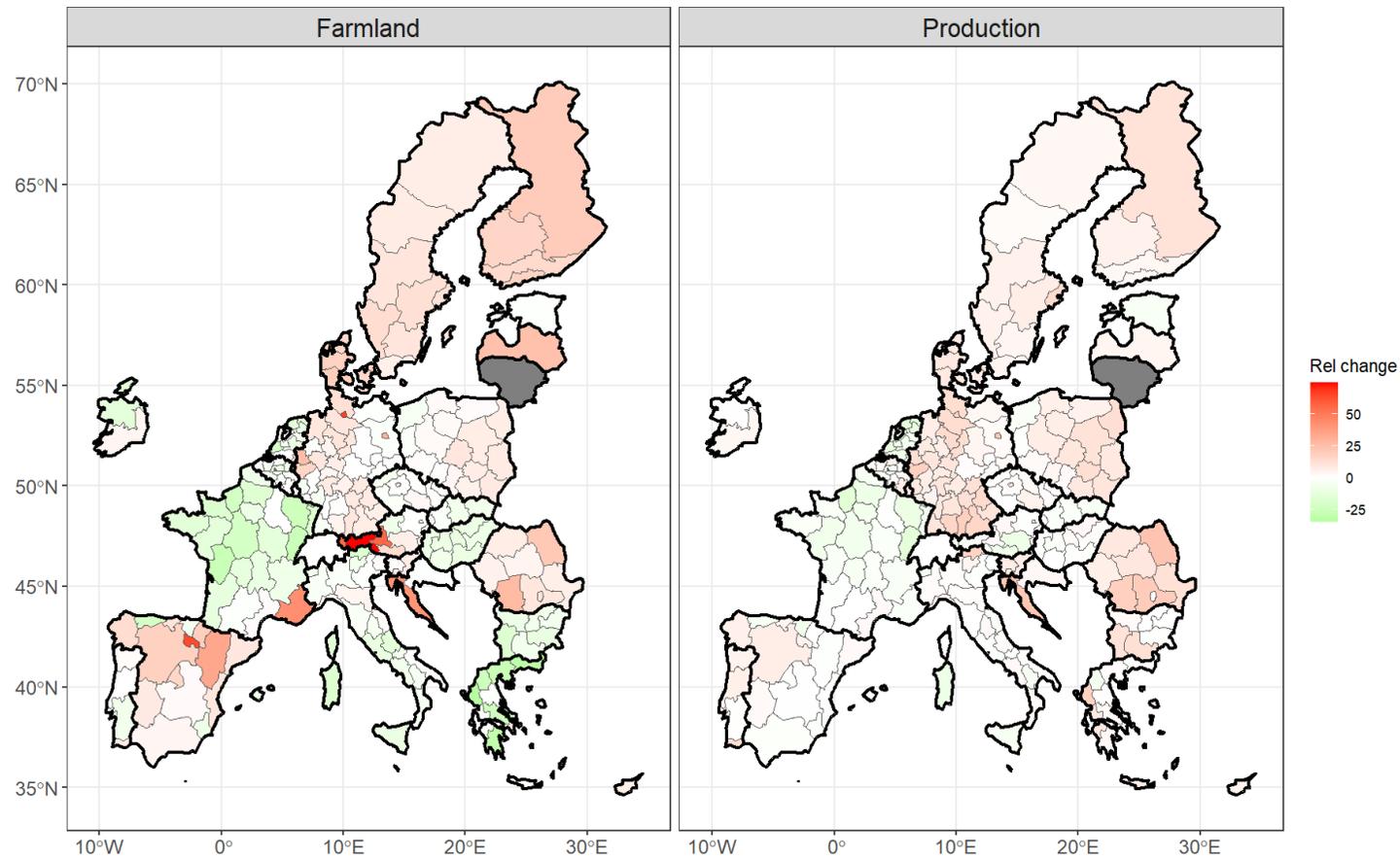

Figure 5: Maps of relative change in Gini index for production and farmland from 2010 to 2020 in Europe

Note: Relative variations of the Gini index are computed as the ratio of the difference between the regional (NUTS-2) Gini index for 2020 and the Gini index for 2010 and the regional Gini index for 2010. Thus, it do coincides with the ratio of the raw (absolute) variation and the Gini in 2010. Relative variations are reported in a percentage scale. Regarding Lithuania (LT), the last available information from Eurostat regards 2016. Thus, Gini index and the corresponding relative variation is not available (grey region).

Figure 4 and Figure 5 shows respectively the raw (i.e., the difference between the Gini in 2020 and the Gini in 2010) and the relative change of the indices between 2010 and 2020. The analysis provides a more granular and informative picture of the European agricultural sector, capturing some dynamics that are clearly lost at aggregate level.

First and foremost, the Gini index for the standard output seems to be always higher than the one for the farmland in all NUTS-2 areas. Although land size is a physical constraint, this result is not totally unexpected, given the fact that a more homogeneous distribution of land does not directly translate into a more homogeneous standard output, since other forces takes place (first and foremost, the specificity of the area with respect of the type of production of farms). Nonetheless, a higher Gini index of the standard output implies that there is substantial higher concentration compared to the actual land distribution. This could be for several reasons, mainly related to the efficiency in the productive process. First, economies of scale could generate higher standard outputs at increasing rates. In other words, a small increase in the land size could translate into a more proportional increase in the standard output, due to an efficiency increase of the land usage. Second, smaller farms could use relatively cheap, and, sometimes old, machinery and equipment, which could potentially decrease the relative output of the farm, while larger farms need to use more efficient machineries and tractors to cover more land. Third, capital investments are typically more efficient for larger-scale companies.

France and the Netherlands seem to be the countries with the lowest levels of Gini index observed, both in terms of standard output and in terms of hectares. In France, between 2010 and 2020 there is a clear decrease in the Gini index with the only exception of the southern part of the country, which records a nil change or a small increase. The decrease is mostly driven by the difference between 2016 and 2020, as shown in Table 1. Moreover, both the Gini index seems, more or less slightly decreasing over time (with more intensity for the hectares concentration), the only exception being the region of the Provence - Alpes – Cote d'Azur in the south, for which the hectares index increased substantially. Nonetheless, regional heterogeneity in terms of land size seems to decrease over time.

Germany and Spain observe very specific patterns. For what concerns the land size, western and southern German NUTS-2 areas show evidently lower Gini index levels with respect to the eastern areas. This difference is constantly verified in the time horizon considered and reflects the historical dualism in development process only partially solved by the reunification in the 90s. There is also some, though mild, evidence of this difference for the standard output. For Spain the story is slightly different. In 2010, northern regions where less concentrated than southern regions in terms of size. However, in the decade under consideration, the northern areas had a substantial increase in the hectares Gini index, showing an increase of concentration, on average, while southern areas had small or nil increase. In addition, this pattern is not observed for the standard output: for Spanish NUTS-2 areas, the Gini index of the standard output is relatively high with minimal variation between areas. It is also worth noticing that that the pattern of southern Spanish regions is very similar to the pattern of Portugal regions, both in terms of hectares and in terms of standard output.

Poland, similarly to Germany, is found to have relevant differences between western and eastern areas. The western area has larger Gini indexes than eastern ones, especially areas on the border with Germany and Czech Republic. Between 2010 and 2020, this gap has been reduced, thanks to the increased concentration of easter regions, although the average standard output Gini index at country level slightly increased in this time window.

The two Scandinavian countries in the sample, Finland and Sweden, present very different level of the Gini index, the first having lower Gini index for both standard output and hectares than the second. In addition, both countries seem to have observed an increase in both Gini index between 2010 and 2020. However, for both countries there are no sizable differences at NUTS-2 level for both the concentration metrics.

The patterns of the Gini indexes for Austria and Italy need careful considerations. Austria has both a relatively low hectares Gini index and a low standard output Gini index, comparable to the averages of western Germany. What is interesting is the time pattern, i.e., how the two indexes have changed between 2010 and 2020: in western Austria (Vorarlberg, Tirol and Salzburg regions) there was a relevant increase in the hectares Gini index, with no relevant change in the standard output concentration; instead, the standard output concentration decreased in Kärnten and Steiermark, regions in which the hectares Gini index slightly increased in the same time horizon. In Italy, the time patterns of the two Gini indexes seem regional-specific. There is a considerable average drop of the hectares Gini indexes in the majority of the regions, with important magnitudes in Sardinia, Lazio and Abruzzo, and, on the other hand, some mild increases, for example in Emilia Romagna and Marche. The standard output Gini index seems to be relatively constant over time, with some regions with mild decreases and others with relatively low increase. Henceforth, for these two countries, the pictures of size and output concentration is still particularly heterogeneous and with a certain degree of regional specificity. Differently from Germany and Poland, well characterized by an East-West dualism, the classical North-South differential in Italy is not patterned as the country shows homogenous values of concentrations in both land and production.

The time patterns of the Gini indexes of Greece are particularly interesting. Starting with an average hectares Gini index similar to the one of the western Germany, it observed the largest absolute drop in our sample in most of the Greek NUTS-2 areas in the 10 years considered. However, in the same period of time it observed different variations of the standard output Gini index, which implies that, while land size became clearly more homogeneous, the standard output became much more concentrated. This is particularly relevant in the Ipeiros area, which actually observed the largest increase in the standard output Gini index and the largest drop in the hectares one.

In the central-eastern part of the Europe, there are some countries with very high level of both Gini indexes, with heterogeneous patterns. Bulgaria had one of the largest hectares Gini index in 2010, which decreased in the next 10 years throughout the national territory, while its standard output Gini index slightly increased, especially in the South Western Region, in the South Central Region and in the North Western Region. In Hungary and Slovakia, a similar pattern is observed, with the only difference being the magnitude of increase of the concentration of standard output, which was nil or very low in both countries.

Ireland, the Netherlands and Belgium have similar levels and patterns of the two Gini indexes, both being relatively low compared to the other countries and with small variations through the decade.

Beside the description of levels and patterns of Gini indexes at country and regional level, it is important to analyze if there are some significant differences in border areas, since those type of areas have obviously similar physical characteristics (e.g. mountain areas between Italy and Austria and the Pyrenees between France and Spain) and also are subject to similar weather conditions, especially in terms of rain, wind and solar irradiation. It is worth noticing that there are some similarities on the agricultural land concentration in some relevant cross-border areas: the Portugal-Spain border; a large part of the central Europe, composed by France, the Netherlands, Belgium, Austria and western Germany; a smaller part of the central Europe, composed by eastern Germany, western Poland and Czech Republic; Hungary and Slovakia. For other areas, the picture is much fragmentated, that common patterns cannot be easily defined. This specificity could be caused by different factors, such as history of conflicts, varying agricultural practices, and land reforms.

The analysis unveils a clear association between the hectares Gini index and the standard output Gini index. Figure 6 and Figure 7 show respectively the linear correlation between the two indexes at year level and the correlation between the change of the two indexes between 2010 and 2020.

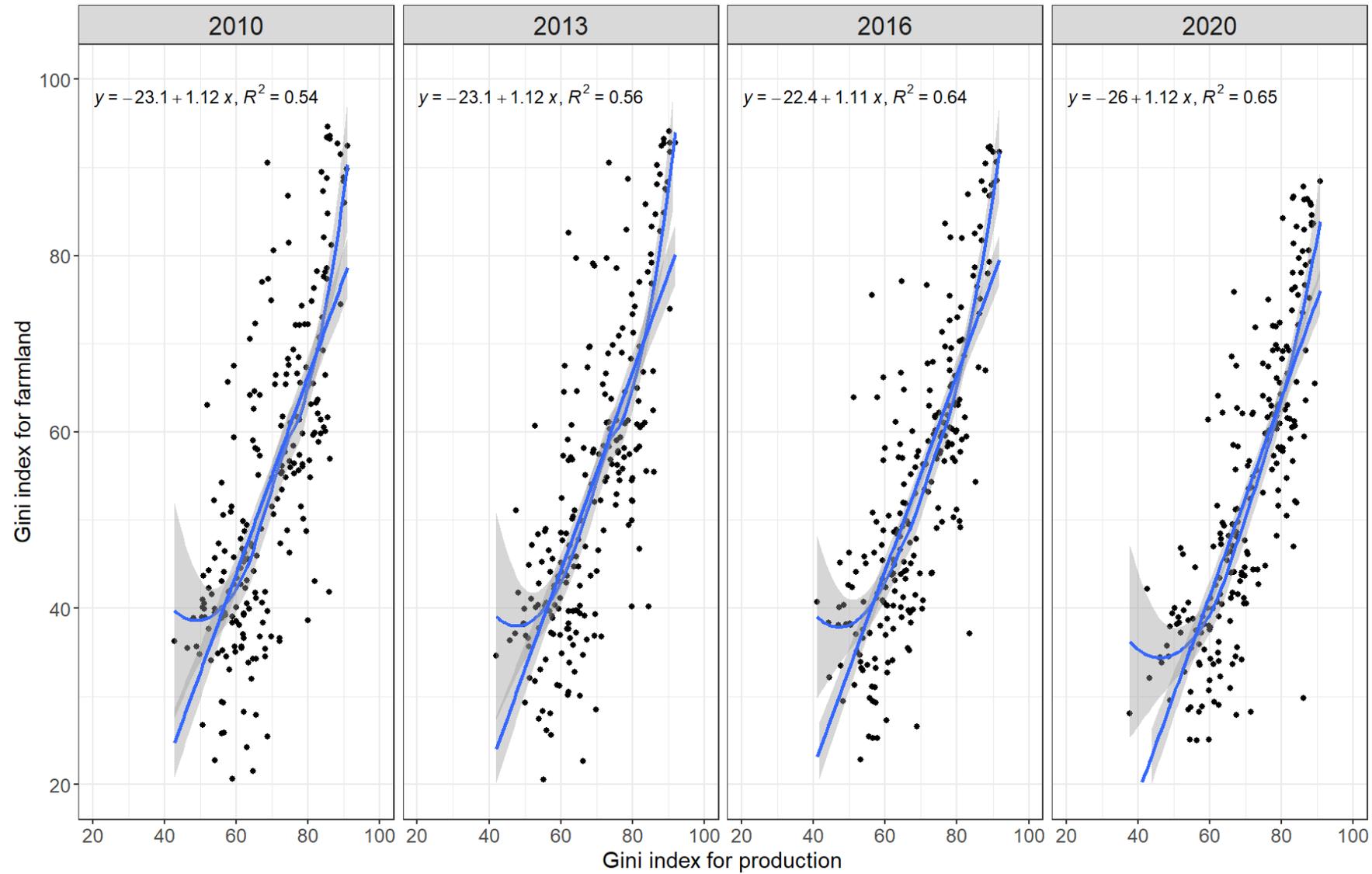

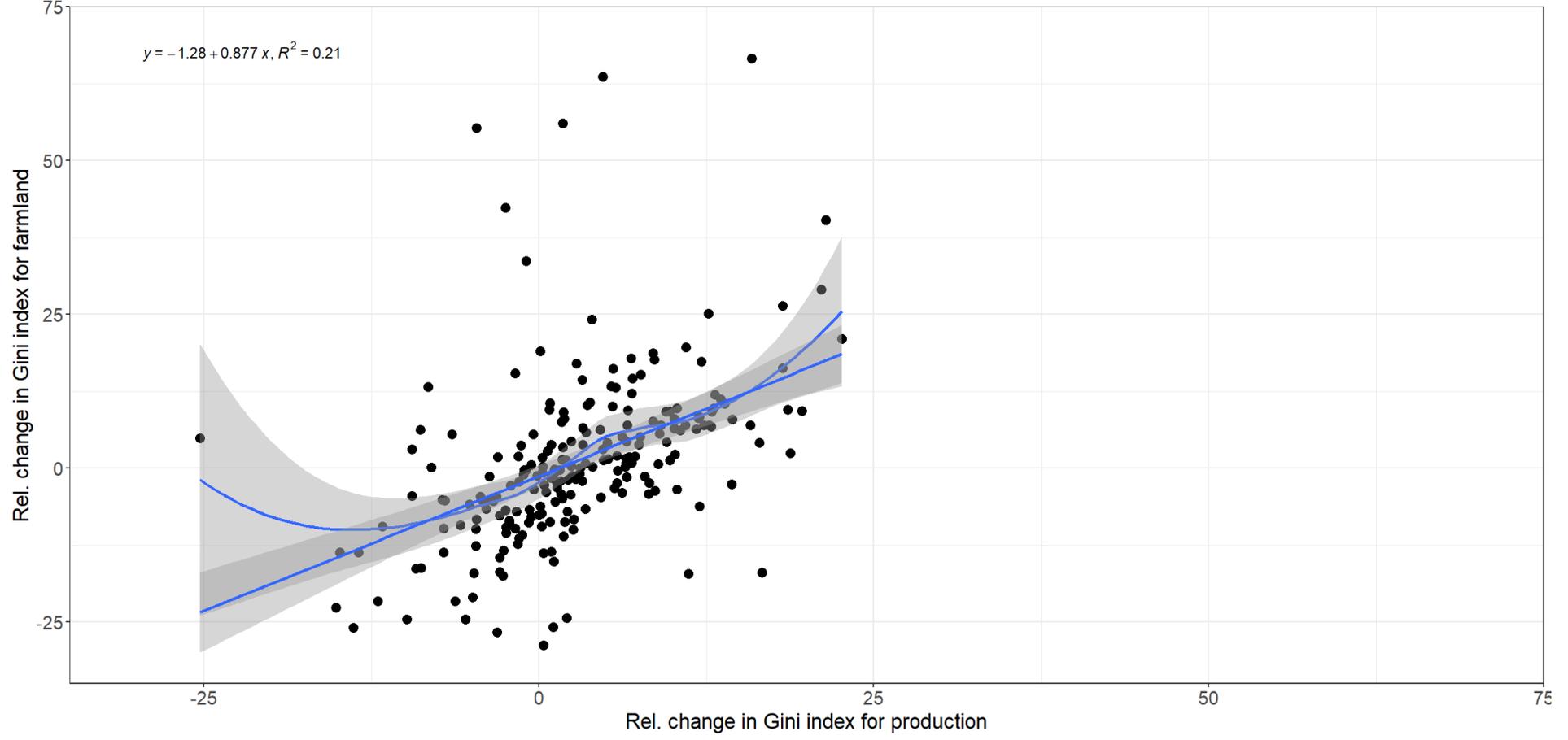

Figure 7: Correlation between relative changes in Gini index for production and farmland

Figure 6 reveals a clear and positive association between hectares and standard output disparity, in all four years considered, with also an increasing $R^2$, meaning that this relation became stronger in the decade. From this figure is also possible to see that, while the range (measured by the distance between the maximum and the minimum value observed) of the Standard Output of the Gini index remained constat throughout the years, the range of the Hectares Gini index shrinked slightly, due to a decrease in the maximum Gini index values and a contemporaneous increase in the minimum levels.

Moving to the spatial dependence analysis, the estimates reported in the Appendix in Table A3 show that, globally (i.e., jointly considering the whole Europe), there is strong positive autocorrelation between regions for both production and farmland concentration. This finding thus supports the hypothesis that adjacent regions tend to register similar levels of statistical concentration, generating clusters of regions with either high and similar or low and similar values. It is worth noting that spatial clusters overcome the national borders of the individual countries, but form contiguous clusters of regions that can be traced back to known historical-political events and processes. Once again, the dichotomy between north-central and southeastern Europe is confirmed. Indeed, from Figure 8 it is obvious that, consistently with Ezcurra et al. (2008), there exists an European dualism in which clusters of regions in southern Europe (i.e., Spain and Italy) and eastern Europe (i.e., Romania and Hungary), register very high concentrations in production and farmland, while Dutch, Belgian and French regions show with high significance a low degree of concentration.

## Figure 8: Spatial autocorrelation using local Moran's index

### Gini index for production

**2010**
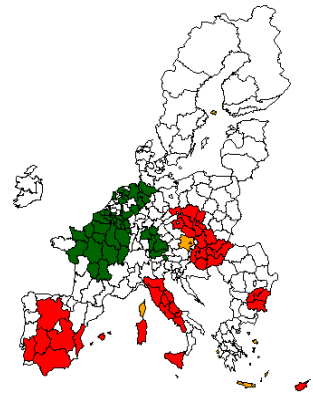

**2013**
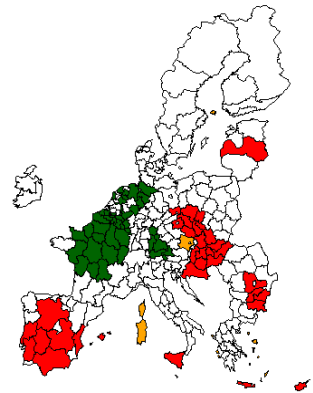

### Gini index for farmland

**2010**
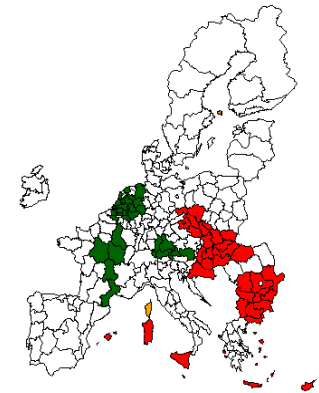

**2013**
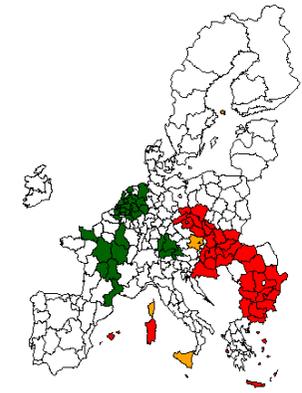

**2016**
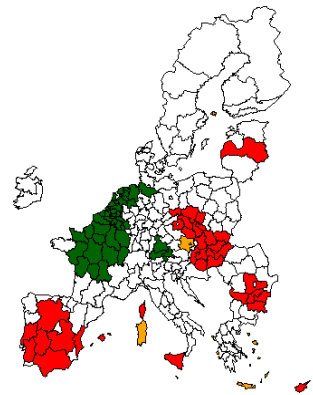

**2020**
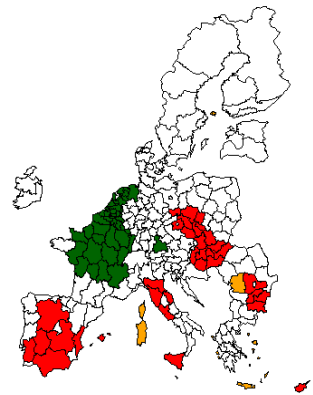

**2016**
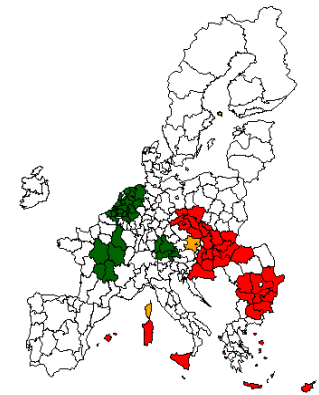

**2020**
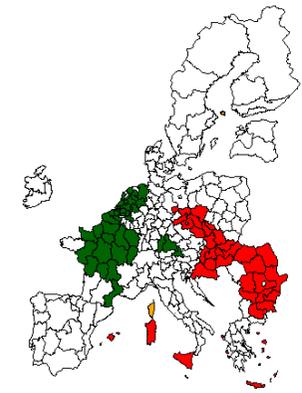

**LISA quarters**  ☐ Not significant  ■ HH  ■ LH  ■ LL

**LISA quarters**  ☐ Not significant  ■ HH  ■ LH  ■ LL

Note: LISA stands for Local Indicator of Spatial Autocorrelation (Anselin, 1995). Regions are grouped into five non-overlapping clusters or quarters, that is, high-high (HH) group (i.e., regions with high concentrations are surrounded by highly-concentrated neighbors); low-low (LL) group (i.e., regions with low concentrations are surrounded by lowly-concentrated neighbors); high-low (HL) group (i.e., regions with high concentrations are surrounded by lowly-concentrated neighbors); low-high (LH) group (i.e., regions with low concentrations are surrounded by highly-concentrated neighbors); non-significant area in which local autocorrelation index is not statistically significant.

A very similar clustering structure is also confirmed in Cerqueti et al. (2024). Indeed, by extending the idea of spatial autocorrelation within a cluster-wise framework, the authors find out that European regions can be clustered into three macro-regions (e.g., Germany, Benelux and North-eastern French regions form a homogeneous cluster) with group-specific determinants of concentrations of agricultural production. The robustness of these results is illustrated in Figure 9, which shows that the spatial autocorrelation remains positive for a large number of spatial lags and even presents increasing values from 2010 to 2020.

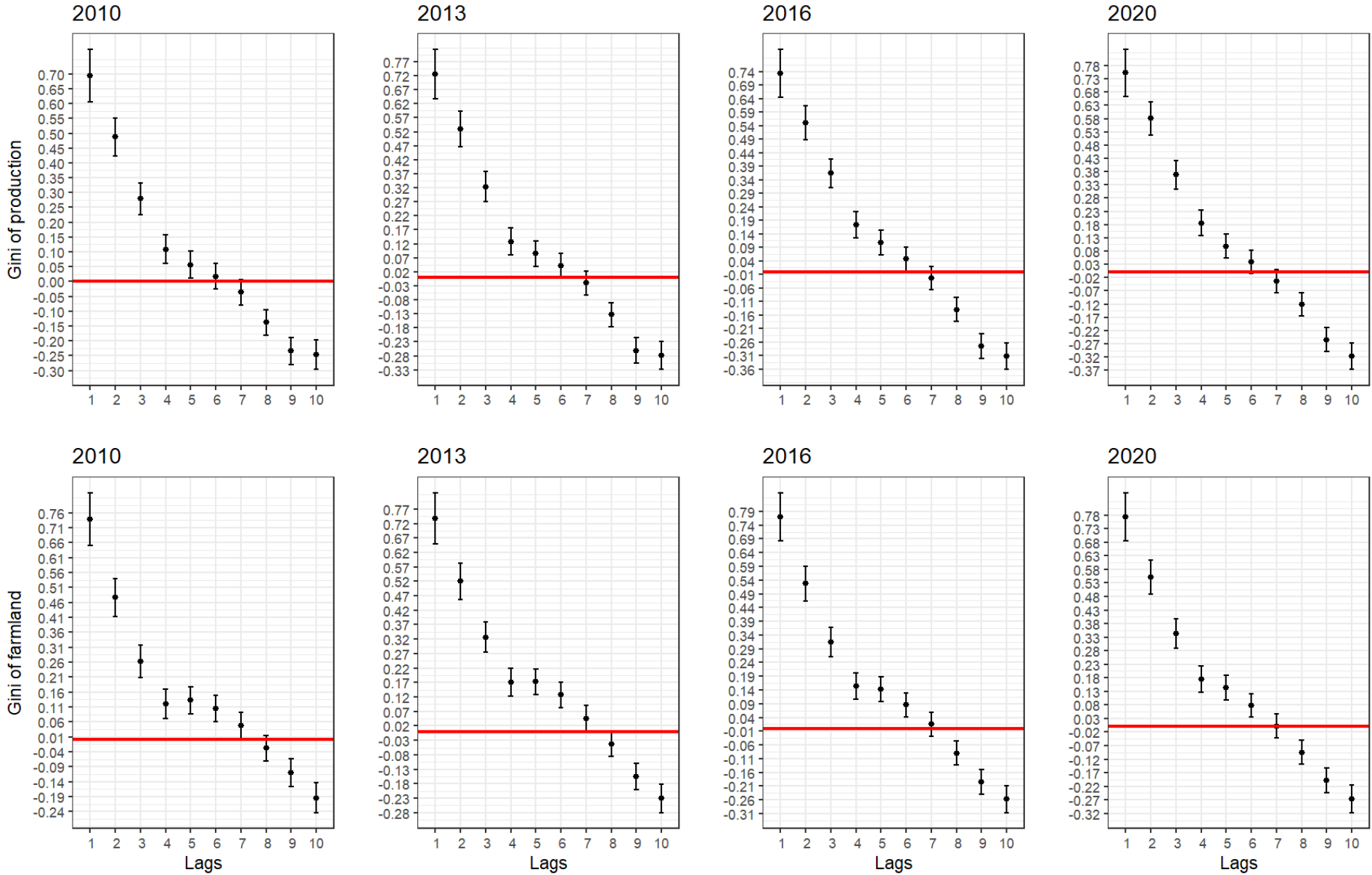

Figure 9: Global Moran's spatial autocorrelogram

## 6. Discussion, Future Work, and conclusive remarks

The research on spatial heterogeneity among European agricultural companies is still underway. Most of the literature takes into account spatial disparity to understand how it affects productivity levels in different regions. Our analysis contributes to this literature by providing a clear mapping of statistical concentration on farmland size and on production at the regional levels for EU-27 countries. In addition, we also provide new evidence regarding the temporal evolution of the concentration of the agricultural sector, focusing on the decade 2010-2020.

This paper provides a fragmented picture of the European agricultural sector, with high country heterogeneity for both farmland and production concentration. The fragmentation of the sector is highlighted by both the within-and-between countries analysis and the study of spatial dependence in the distribution of the Gini index. Spatial heterogeneities are evident both within and between countries, with noticeable clusters of neighboring highly-concentrated regions opposed to clusters of adjacent regions with significantly lower concentrations. Such clusters transcend national boundaries and often correspond to areas with known historical and political processes (e.g., former Soviet bloc countries). We also show that there is a positive association between the two disparity measures, implying that both economies of scale and economies of scope are relevant in this sector. Several time patterns clearly emerge, but highly country and intra-country specific. We are also able to define limited cross-country areas with similar patterns of land and output concentrations.

According to our findings, at aggregate level, the more equal distribution (or less unequal) distribution of land and standard output is found in the area delimited at west by Austria and western Germany, at east by France, at north by the Netherlands and at south by southern France and Austria. Instead, Central (from eastern Germany) and Eastern Europe seems to have the more unequal distribution of both the indicators, especially Czech Republic, Hungary, Romania, Bulgaria and Slovakia. At the end of the XXth century, Central and Easter countries had been under significant changes in their agricultural sector, especially the approach of land distribution and some escape from rural areas. Specifically, as van Vliet et al. (2015) argue, land use has been significantly impacted by the shift to post-socialism, since under socialism, most of land were collectivized under some optimization schemes run at central level. Afterwards, the de-collectivization of lands and the return to private holdings had significant impacts on land usage and on the escape from very poor rural areas. This could explain why still nowadays in those countries we have very high unequal distribution of land, as large firm/companies could have found some monopoly power and also acquired many small firms. Between 1990 and 2006, Levers et al. (2018) show that low-intensity and de-intensifying land systems dominated in Europe's east, in clear contradiction with the dynamics of western Europe. Also, some countries belonging to the southern part of Europe (namely, Spain, Italy, Portugal and Croatia) seem to have, on aggregate, similar level of disparities, slightly higher than the ones registered for France and western Germany or Benelux countries. Mediterranean European areas have different characteristics in terms of soil, degradation related to the climate change and other factors such as farmland abandonment (which is a common factor in Europe) with respect to central Europe (e.g., see García-Ruiz & Lana-Renault, 2011; Malek et al., 2018).

Our analysis shows also some interesting results within countries. The first result, already explained in the previous section, is the difference between Western and Easter Germany, that is clearly related to the historical dualism. In Spain we find some clear differences in terms of hectares disparity between northern and southern regions. This could be caused by difference in the climate, geography, and soil quality, which have consequences on the type of farming and the need for smaller or bigger firm in some cases. Also land abandonment has been different between rural areas of regions like the Pyrenees with respect to other regions (García-Ruiz, 2010). On the contrary, we do not find evidence of a clear difference between the land distribution of north vs south of Italy. Even though, historically, differences in land management were present (e.g., see Corti et al., 2013 for a comprehensive historical picture of the Italian land management), the picture that emerges from our analysis is more fragmented, with some differences between regions but not resulting in a clear pattern.

Eventually, it is worth noting that the development of rural areas represents one of the main challenges of the EU's PAC strategy. The capacity of the Member States and the European Union to ensure a process of revitalization of rural areas, reversing the current process of desertification, allows for greater social and territorial cohesion, but is also a response to the major climate challenges. The agroforestry mosaic, being one of the main eco-schemes[8] provided in the current CAP regulations, requires understanding the advantages of ensuring the viability of a scale of production that has been relegated to secondary importance for decades. Small and medium-sized farms recognize the productive potential of vast abandoned areas, which can be allocated to forestry, plant or animal production, feeding short marketing circuits, boosting local economies and ensuring prevention of the fires that have devastated much of Europe's forests.

Summing up, regarding our first research question, our analysis does not fully unveil common trajectories on the level of land and production disparities in Europe between 2010 and 2020. Instead, delving into the second research question, we find a clear positive association between the two measures of market concentration. This could be a result of the significant economies of scale present in the agricultural sector.

---

[8] Eco-schemes provide support for farmers who observe agricultural practices beneficial for the environment and climate. It is a measure to reward and incentivize farmers for acting towards a more sustainable farm and land management with the objective to maintain public goods.

This work has some straightforward follow-ups. First and foremost, the paper could be extended by considering more detailed classifications of farms by splitting the producers between organic and non-organic production, as well as between crops and livestock farmers. This would enrich the overall picture and help us understand whether the large heterogeneity in the output could be driven by those type of farms. Second, it could be interesting to understand whether the heterogeneity in the statistical concentration of standard output is driven by different land productivity or large-scale investments. Third, given the increase importance of sustainable agronomic practices, it could be interesting to see whether increases or decreases of land and output concentrations are connected to changes in the pollution level at regional level, especially considering the role of ammonia emissions in producing fine particulate matters (Otto et al., 2024). This could help policymakers to understand whether a more equal land distribution could have an impact also in reducing GHG emissions. Lastly, the link between farmland and production concentration on consumer behavior could be assessed. This is relevant in context where very large companies increase their market share and reach some monopolistic power.

# Declarations

- <u>Conflict of interest/Competing interests</u>: The authors have no competing interests to declare that are relevant to the content of this article.
- <u>Funding</u>: The authors did not receive support from any organization for the submitted work.
- <u>Data availability and codes</u>: All results presented in this paper can be reproduced using the R software. The codes were developed entirely by the authors. Fore reproducibility purposes, all the scripts and the data are made public through the following GitHub folder https://github.com/PaoloMaranzano/SB_PM_MV_AgroMarketConc.git
- <u>Figures</u>: All images included in the paper were created by the authors and do not require any publication permission.

# Acknowledgements

This is a preprint of the following chapter: *Boccaletti, S., Maranzano, P. and Viegas, M.*, **Inequality and Concentration in Farmland Production and Size: Regional Analysis for the European Union from 2010 to 2020**, published in Global Perspectives on Climate Change, Inequality, and Multinational Corporations, edited by João Bento & Miguel Torres, 2024, Springer Nature Switzerland AG reproduced with permission of Springer Nature Switzerland AG.

# Appendix

**Table A1: Average Gini index by Country and Year**

|         | Gini index for farmland |       |       |       | Gini index for production |       |       |       |
|---------|-------|-------|-------|-------|-------|-------|-------|-------|
| Country | 2010  | 2013  | 2016  | 2020  | 2010  | 2013  | 2016  | 2020  |
| AT      | 34.42 | 40.01 | 43.82 | 39.60 | 63.59 | 62.78 | 62.97 | 62.61 |
| BE      | 37.17 | 35.42 | 36.28 | 36.12 | 55.26 | 55.35 | 53.19 | 54.63 |
| BG      | 92.02 | 91.95 | 89.62 | 83.37 | 81.02 | 84.67 | 85.89 | 83.93 |
| CY      | 61.64 | 62.45 | 66.96 | 65.52 | 85.46 | 85.74 | 87.88 | 89.37 |
| CZ      | 77.27 | 76.63 | 78.10 | 77.38 | 83.30 | 81.44 | 84.27 | 85.54 |
| DE      | 45.80 | 46.33 | 47.23 | 48.10 | 60.55 | 61.23 | 62.47 | 66.18 |
| DK      | 54.51 | 56.62 | 58.19 | 63.59 | 73.11 | 73.12 | 74.53 | 79.53 |
| EE      | 74.44 | 73.91 | 77.92 | 73.48 | 89.15 | 90.38 | 88.68 | 85.88 |
| EL      | 64.14 | 65.71 | 54.92 | 53.73 | 64.91 | 66.42 | 64.95 | 66.72 |
| ES      | 51.67 | 52.30 | 56.35 | 55.77 | 77.82 | 78.37 | 78.08 | 78.14 |
| FI      | 37.22 | 36.48 | 38.16 | 40.75 | 65.86 | 64.27 | 66.06 | 68.44 |
| FR      | 38.64 | 38.06 | 40.64 | 33.61 | 57.80 | 56.04 | 57.16 | 55.24 |
| HR      | 56.26 | 73.98 | 62.95 | 64.77 | 63.38 | 66.93 | 65.84 | 69.93 |
| HU      | 89.56 | 89.94 | 88.78 | 81.19 | 86.71 | 88.65 | 89.18 | 86.71 |
| IE      | 36.94 | 37.17 | 37.50 | 35.15 | 67.17 | 65.58 | 66.61 | 67.85 |
| IT      | 62.71 | 56.59 | 59.70 | 59.44 | 77.47 | 72.21 | 72.11 | 77.37 |
| LT      | 65.53 | 68.29 | 70.46 | NA    | 77.92 | 79.67 | 81.51 | NA    |
| LU      | 43.64 | 44.89 | 44.07 | 44.64 | 50.80 | 51.16 | 51.77 | 55.97 |
| LV      | 60.48 | 66.88 | 67.31 | 75.13 | 84.44 | 85.79 | 86.06 | 87.79 |
| MT      | 41.86 | 40.17 | 37.11 | 29.83 | 85.81 | 84.53 | 83.57 | 86.13 |
| NL      | 34.55 | 36.19 | 31.44 | 32.35 | 60.06 | 59.70 | 53.53 | 53.83 |
| PL      | 54.08 | 54.86 | 54.86 | 55.61 | 70.38 | 70.76 | 72.36 | 74.48 |
| PT      | 61.49 | 61.75 | 62.36 | 61.21 | 79.32 | 80.24 | 80.65 | 83.29 |
| RO      | 68.63 | 71.63 | 67.98 | 75.89 | 69.34 | 72.42 | 71.16 | 77.33 |
| SE      | 56.56 | 58.04 | 60.29 | 61.97 | 75.70 | 78.79 | 78.39 | 79.74 |
| SI      | 47.47 | 48.46 | 47.30 | 49.39 | 66.36 | 66.22 | 66.80 | 70.85 |
| SK      | 88.28 | 87.06 | 87.55 | 85.25 | 90.22 | 89.32 | 89.94 | 88.45 |

Note: Average Gini index by country and year is computed as the arithmetic mean of the available NUTS-2 values. Regarding Lithuania (LT), the last available information from Eurostat regards 2016. Thus, Gini index and the corresponding descriptive statistics for 2020 are not available (NA).

## Table A2: Standard deviation of the Gini index by Country and Year

| Country | Gini index for farmland | | | | Gini index for production | | | |
|---|---|---|---|---|---|---|---|---|
| | 2010 | 2013 | 2016 | 2020 | 2010 | 2013 | 2016 | 2020 |
| AT | 10.16 | 11.91 | 8.62 | 5.71 | 5.84 | 6.51 | 6.45 | 5.57 |
| BE | 5.79 | 6.75 | 5.90 | 5.47 | 5.48 | 6.99 | 5.06 | 6.55 |
| BG | 2.91 | 1.97 | 3.59 | 4.72 | 7.55 | 6.93 | 5.03 | 3.95 |
| CY | / | / | / | / | / | / | / | / |
| CZ | 3.64 | 7.75 | 3.15 | 4.71 | 2.69 | 8.23 | 3.58 | 2.87 |
| DE | 12.31 | 11.70 | 11.31 | 10.64 | 8.08 | 6.92 | 7.41 | 6.27 |
| DK | 2.66 | 3.29 | 3.18 | 2.41 | 1.69 | 2.28 | 2.54 | 2.81 |
| EE | / | / | / | / | / | / | / | / |
| EL | 8.05 | 9.80 | 8.03 | 6.94 | 6.75 | 4.50 | 4.99 | 7.81 |
| ES | 11.44 | 10.50 | 7.67 | 10.54 | 4.51 | 4.45 | 3.78 | 4.51 |
| FI | 1.93 | 1.00 | 3.13 | 4.14 | 3.27 | 3.84 | 3.72 | 4.60 |
| FR | 8.17 | 8.18 | 7.88 | 5.64 | 5.64 | 6.06 | 7.10 | 6.85 |
| HR | 16.98 | 12.11 | 6.79 | 3.74 | 15.56 | 6.77 | 8.49 | 8.97 |
| HU | 2.93 | 2.92 | 2.98 | 3.23 | 2.76 | 2.22 | 2.09 | 1.90 |
| IE | 3.78 | 3.50 | 3.20 | 3.47 | 2.30 | 1.05 | 0.06 | 1.47 |
| IT | 7.10 | 7.56 | 7.15 | 6.69 | 7.63 | 9.36 | 8.64 | 6.61 |
| LT | / | / | / | / | / | / | / | / |
| LU | / | / | / | / | / | / | / | / |
| LV | / | / | / | / | / | / | / | / |
| MT | / | / | / | / | / | / | / | / |
| NL | 5.24 | 4.70 | 4.01 | 4.12 | 4.65 | 4.69 | 4.83 | 6.68 |
| PL | 12.25 | 11.35 | 11.02 | 10.65 | 7.95 | 6.96 | 6.32 | 5.36 |
| PT | 11.81 | 10.42 | 12.69 | 9.17 | 6.18 | 4.58 | 6.24 | 4.89 |
| RO | 10.51 | 11.06 | 11.10 | 9.35 | 9.06 | 8.04 | 7.93 | 6.40 |
| SE | 1.76 | 1.58 | 1.30 | 1.26 | 3.33 | 2.61 | 2.25 | 1.80 |
| SI | 2.15 | 2.11 | 3.10 | 4.27 | 0.80 | 0.80 | 0.43 | 0.70 |
| SK | 1.63 | 1.51 | 0.83 | 2.46 | 0.40 | 0.46 | 0.88 | 1.83 |

Note: Standard deviation of Gini index by country per year is computed as the standard deviation of the available NUTS-2 values. For Cyprus (CY), Estonia (EE), Lithuania (LT), Luxembourg (LU), Latvia (LV), and Malta (MT) the only available information from Eurostat are at the national (NUTS-0) level. Thus, no infra-country variability can be computed as, for each year, only a single numeric value is provided.

**Table A3: Moran's I statistics and p-values for the Gini index on production and farmland from 2010 to 2020**

| Gini index | Statistics | 2010 | 2013 | 2016 | 2020 |
|---|---|---|---|---|---|
| **Production** | Moran's I estimate | 0.6457 | 0.6731 | 0.6895 | 0.7218 |
| | P-value parametric | 0.0000 | 0.0000 | 0.0000 | 0.0000 |
| | P-value Monte Carlo | 0.001 | 0.001 | 0.001 | 0.001 |
| **Farmland** | Moran's I estimate | 0.698 | 0.6908 | 0.7216 | 0.7486 |
| | P-value parametric | 0.0000 | 0.0000 | 0.0000 | 0.0000 |
| | P-value Monte Carlo | 0.001 | 0.001 | 0.001 | 0.001 |

Note: Moran's I estimate refers to the 1-lag Moran's statistics for spatial autocorrelation of the observations under the null hypothesis that Gini indices are randomly distributed across European regions following a completely random (non-spatial) process. Parametric p-values for Moran's I statistics are computed assuming a Gaussian distribution, while Monte Carlo p-values are computed using the empirical distribution of the test statistics from N=1000 independent replications.